\documentclass[preprint,showpacs,preprintnumbers,amsmath,amssymb,nofootinbib]{revtex4}
\usepackage{booktabs}
\usepackage{mathrsfs}
\usepackage{epsfig}
\usepackage{graphicx}
\usepackage{dcolumn}
\usepackage{bm}
\usepackage{amsmath}
\usepackage{slashed}

\let\jnfont=\rm
\def\NPB#1,{{\jnfont Nucl.\ Phys.\ B }{\bf #1},}
\def\PLB#1,{{\jnfont Phys.\ Lett.\ B }{\bf #1},}
\def\EPJC#1,{{\jnfont Eur.\ Phys.\ Jour.\ C }{\bf #1},}
\def\PRD#1,{{\jnfont Phys.\ Rev.\ D }{\bf #1},}
\def\PRL#1,{{\jnfont Phys.\ Rev.\ Lett.\ }{\bf #1},}
\def\MPLA#1,{{\jnfont Mod.\ Phys.\ Lett.\ A }{\bf #1},}
\def\JPG#1,{{\jnfont J.\ Phys.\ G}{\bf #1},}
\def\CTP#1,{{\jnfont Commun.\ Theor.\ Phys.\ }{\bf #1},}
\def\ZPC#1,{{\jnfont Z.\ Phys.\ C }{\bf #1},}
\def\JHEP#1,{{\jnfont JHEP \ }{\bf #1},}
\def\Rv{\not{\hbox{\kern-1pt $R$}}}
\def\p{\not{\hbox{\kern-3pt $p$}}}

\begin{document}
\preprint{\parbox{1.2in}{\noindent arXiv:1101.4456 }}

\title{Top quark forward-backward asymmetry, FCNC decays and like-sign pair production as a joint probe of new physics}

\author{Junjie Cao$^1$, Lin Wang$^{1,2}$, Lei Wu$^2$, Jin Min Yang$^2$
        \\~ \vspace*{-0.3cm} }
\affiliation{
$^1$ College of Physics and Information Engineering, Henan Normal University,
     Xinxiang 453007, China\\
$^2$ Key Laboratory of Frontiers in Theoretical Physics,
     Institute of Theoretical Physics,
     Academia Sinica, Beijing 100190, China
     \vspace*{1.5cm}}

\begin{abstract}
The anomaly of the top quark forward-backward asymmetry $A^{t}_{FB}$
observed at the Tevatron can be explained by the $t$-channel
exchange of a neutral gauge boson ($Z'$) which has sizable flavor
changing coupling for top and up quarks. This gauge boson can also
induce the top quark flavor-changing neutral-current (FCNC) decays
and the like-sign top pair production at the LHC. In this work we
focus on two models which predict such a $Z'$, namely the left-right
model and the $\mathrm{U}(1)_X$ model, to investigate the correlated
effects on $A^{t}_{FB}$, the FCNC decays $t \to u V$
($V=g,Z,\gamma$) and the like-sign top pair production at the LHC.
We also pay special attention to the most recently measured
$A^{t}_{FB}$ in the large top pair invariant mass region. We find
that under the current experimental constraints both models can
alleviate the deviation of $A^{t}_{FB}$ and, meanwhile, enhance the
like-sign top pair production to the detectable level of the LHC. We
also find that the two models give different predictions for the
observables and their correlations, and thus they may even be
distinguished by jointly studying these top quark observables.
\end{abstract}


\maketitle

\section{INTRODUCTION}
In the Standard Model (SM) the top quark is the only fermion with a
mass at the electroweak symmetry breaking scale and hence is
speculated to be a window on new physics beyond the SM
\cite{top-review-th}. So far the Tevatron has measured some
properties of the top quark and found good agreements with the SM
predictions except for the forward-backward asymmetry $A^{t}_{FB}$,
which shows a $2\sigma$ deviation from the SM expectation
\cite{top-afb-exp1}. Although the latest analysis based on the $5.3
fb^{-1}$ luminosity reduced the deviation to about $1.8\sigma$, it
indicated that the forward-backward asymmetry depends on the top
pair invariant mass $M_{t\bar{t}}$ and for $M_{t\bar{t}}\geq 450$
GeV the deviation is enlarged to $3.4\sigma$ \cite{top-afb-exp2}. So
far various new physics schemes have been proposed to explain such a
deviation \cite{top-afb-th1,top-afb-th2,top-afb-th3,top-afb-th4},
among which one attractive way is the $t$-channel exchange of a
neutral gauge boson $Z^{\prime}$ which has sizable FCNC coupling for
top and up quarks.

These $Z^{\prime}$-models are especially interesting because, in
addition to contributing to $A_{FB}^{t}$, they can also induce the
top quark FCNC decays and the like-sign top pair production at the
LHC. Due to their suppressed rates in the SM \cite{fcnc-sm} and the
rather clean backgrounds \cite{like-sign top-th1,like-sign
top-th2,like-sign top-th3,like-sign top-th4}, these FCNC decays and
like-sign top pair production can be a good further test of the FCNC
$Z^{'}$ models for explaination of $A^{t}_{FB}$. On the other hand,
the forward-backward asymmetry, the FCNC decays and the like-sign
top pair production are correlated with each other and such
correlations are model-dependent and thus can help to distinguish
different models at the LHC. In this work we concentrate on two such
$Z^{\prime}$-models, i.e. the left-right model and the
$\mathrm{U}(1)_X$ model, to study the correlated effects of top
quark forward-backward asymmetry at the Tevatron, the FCNC decays
$t\to u V$ ($V=g,Z,\gamma$) and the like-sign top pair production at
the LHC.

This work is organized as follows.  In Sec. II we briefly describe
the two models and present the calculation of the observables.
In Sec. III some numerical results are presented.
Finally, we draw conclusions in Sec. IV.

\section{Models and Calculations}

\begin{figure}[htb]
\epsfig{file=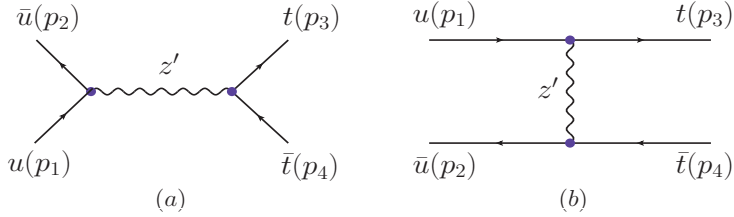,width=10cm,height=3cm} \vspace{-0.5cm}
\caption{Feynman diagrams contributing to $t\bar{t}$ production.}
\label{fig1}
\end{figure}

In extensions of the SM with some extra gauge symmetry, a new
neutral gauge boson called $Z^{\prime}$ is often predicted with
family universal or non-universal couplings to fermions. For a
$Z^\prime$ with family universal couplings, it contributes to the
$t\bar{t}$ production only via the $s$-channel exchange shown in
Fig.\ref{fig1}(a). Such $s$-channel contribution does not interfere
with the dominant QCD amplitude and thus can not sizably enhance
$A^t_{FB}$. However, the situation is quite different for a
$Z^\prime$ with non-universal couplings where the FCNC interaction
$Z^{\prime} \bar{t}_R u_R$ may arise without violating the
constraints from flavor physics \cite{top-afb-th3}. In this case,
$Z^\prime$ can contribute to $t\bar{t}$ production via the
$t$-channel exchange shown in Fig.\ref{fig1}(b), which, as pointed
out in \cite{top-afb-th3}, can interfere with the QCD amplitude to
enhance $A^t_{FB}$ significantly while alter the cross section
mildly. Since we attempt to explain the deviation of $A^t_{FB}$ by
new physics effects, we in this work consider two models with
non-universal $Z^\prime$ couplings, which are called model-I and
model-II, respectively.

Model-I extends the SM by a $U(1)_X$ gauge symmetry under which the
three generations of the right-handed up-type quarks are charged as
$(-1+\epsilon_U, \epsilon_U, 1 + \epsilon_U)$ \cite{top-afb-th3}.
Therefore the new neutral gauge boson $Z^{'}$ predicted in this
model only couples to the right-handed quarks, and can safely escape
the constraints from Drell-Yan measurement at the Tevatron and
LHC\cite{drell-yan}. Meanwhile, in this framework, the Yukawa
couplings may be generated by the Froggatt-Nielsen type mechanism
\cite{Froggatt} and the chiral gauge anomalies can be avoided by
introducing some extra fermions. The Lagrangian relevant to our
discussion is \cite{top-afb-th3}
\begin{eqnarray}
{\cal L}_I&=&
g_{x}\bar{u}\gamma^{\mu}P_{R}tZ^{\prime}_{\mu} + \sum_{i=1}^3
\epsilon_{U}g_{x}\bar{u_{i}}\gamma^{\mu}P_{R}u_{i}Z^{\prime}_{\mu}
\end{eqnarray}
where $g_x$ and  $\epsilon_{U}$ are dimensionless parameters, and
$i$ is the generation index. Recently, it is found
that this model can explain the deviation of $A^t_{FB}$ and meanwhile
satisfy other Tevatron measurements in the parameter region:
$120{\rm GeV} < m_Z^\prime < 170 {\rm GeV}$, $\alpha_x = g_x^2/(4\pi) < 0.05$
and $\epsilon_U \leq O(1)$,
which is obtained by the following consideration \cite{top-afb-th3}:
\begin{itemize}
\item If $Z^\prime$ is heavier than top quark, it
will decay dominantly to $t \bar{u}$ or $u \bar{t}$ and
consequently give excessive like-sign top quark events
through the processes $u u \to tt $, $u g \to t Z^\prime \to t t
\bar{u}$ and $ u \bar{u} \to Z^\prime Z^\prime \to t \bar{u} t \bar{u}$.

\item If $Z^\prime$ is much lighter than top quark, the exotic decay
$t \to u Z^\prime$ is open.
For $m_{Z^\prime} < 120 {\rm ~GeV}$ with  $\alpha_X=0.01$,
the decay rate will exceed $10\%$, which can cause a
tension between the dileptonic and hadronic
channels for the  $t\bar{t}$ production at the Tevatron
because such a light $Z^\prime$ will decay into light quarks.

\item The presence of a small $\epsilon_U$ is  necessary
to make the model phenomenologically viable.
If $\epsilon_U=0$, the only decay mode of $Z^\prime$ is $Z^\prime \to t \bar{u}$,
and then both $u\bar{u} \to Z^\prime Z^\prime$ and $u g \to t Z^\prime$
can give the very similar like-sign top pair signal, which is
strongly constrained by the Tevatron experiment.
A non-zero $\epsilon_U$ can avoid this conflict by allowing $Z^\prime$ to
decay into $u\bar{u}$.
On the other hand, $\epsilon_U$ can not be too large
because it can enhance the rates of both $p\bar{p} \to
Z^\prime \to dijet$ and the loop induced decay $t \to u g$,
which have been constrained by the measurements at the Tevatron.
\end{itemize}

Model-II is a special
left-right symmetric model called the third-generation enhanced
left-right model, which is based on the gauge group $SU(3)_C \times
SU(2)_L\times SU(2)_R \times U(1)_{B-L}$ with gauge couplings $g_3$,
$g_L$, $g_R$ and $g$ respectively\cite{hexg}. The key feature of
this model is that the gauge bosons of the $SU(2)_R$ group couple only to
the third-generation fermions (the third generation is
specially treated) \cite{hexg,top-afb-th4}.
The gauge interactions relevant to our study are given by
\small
\begin{eqnarray}
{\cal L}^{Q}_{II}&=& -{g_L\over 2 \cos\theta_W} \bar q \gamma^\mu
(g_V - g_A \gamma_5) q (\cos\xi_Z Z_\mu - \sin\xi_Z Z^\prime_\mu)
\nonumber\\
&&+ {g_Y\over 2} \tan\theta_R ({1\over 3} \bar q_L \gamma^\mu q_L+
{4\over 3} \bar u_{Ri} \gamma^\mu u_{Ri} -{2\over 3} \bar
d_{Ri}\gamma^\mu d_{Ri})
(\sin\xi_Z Z_\mu + \cos\xi_Z Z^\prime_\mu)\nonumber\\
&&- {g_Y\over 2} (\tan\theta_R + \cot\theta_R) ( \bar u_{Ri}
\gamma^\mu V^{u*}_{Rti} V^{u}_{Rtj}u_{Rj} - \bar d_{Ri} \gamma^\mu
V^{d*}_{Rbi} V^{d}_{Rbj} d_{Rj}) (\sin\xi_Z Z_\mu + \cos\xi_Z
Z^\prime_\mu) \label{neucoup}
\end{eqnarray}
where $\tan \theta_R = g/g_R$, $g_Y = g \cos\theta_R = g_R
\sin\theta_R$, $\xi_{Z}$ is the mixing angle between $Z_R$ and
$Z_0$, $V^{u,d}_{Rij}$ are the unitary matrices which rotate the
right-handed quarks $u_{Ri}$ and $d_{Ri}$ from interaction basis to
mass eigenstates and the repeated generation indices $i$ and $j$ are
summed. Similar to model-I, a sizable $u_R-t_R$ mixing with other
flavor mixings suppressed is allowed by the low energy flavor
physics \cite{top-afb-th3}. Such a sizable $u_R-t_R$ mixing
can lead to a rather strong $Z^\prime \bar{t}_R u_R$ interaction
with the condition $g_R \gg g_Y$.

Similarly, the interaction to leptons are given as
follows\cite{hexg}:
\begin{eqnarray}
{\cal L}^{L}_{II}&=& - {g_L\over 2 \cos\theta_W} \bar \ell
\gamma^\mu (g_V - g_A\gamma_5) \ell (\cos\xi_Z Z_\mu - \sin\xi_Z
Z^\prime_\mu)
\nonumber\\
&+& {g_Y\over 2} \tan\theta_R (- \bar \ell_L \gamma^\mu \ell_L -2
\bar E_{Ri} \gamma^\mu E_{Ri})
(\sin\xi_Z Z_\mu + \cos\xi_Z Z^\prime_\mu)\nonumber\\
&-& {g_Y\over 2} (\tan\theta_R + \cot\theta_R) ( \bar \nu_{R \tau}
\gamma^\mu \nu_{R \tau} - \bar \tau_R \gamma^\mu \tau_R) (\sin\xi_Z
Z_\mu + \cos\xi_Z Z^\prime_\mu). \label{coupslep}
\end{eqnarray}

The constraints on model-II were found \cite{hexg} to be: $\cot
\theta_R \leq 20$ from the requirement of perturbativity,
$M_{Z^\prime} \gtrsim 460$ GeV for $\cot \theta_R \ge 10$ from the
global fit of the LEP data (especially $R_b$). Among the oblique
parameters, $T$ gives the most stringent constraint, which roughly
requires $\xi_Z M_{Z^\prime}/(500{\rm GeV}) < 0.01 $ at $3 \sigma$
level. Note that the constraints on $m_{Z^\prime}$ from the CDF
search for new resonant states or from the global fitting of the
electroweak precision data are not applicable here since they
usually assume a $Z^\prime$ with family universal couplings. From
the Eqs.(2) and (3), we can see that the flavor-conserving
interactions between the dominantly right-handed $Z^{'}$ and the
first two generation fermions are suppressed by small $\tan\theta$
for the chosen parameters in our calculation, and thus make a
negligible contribution to the process $pp \to Z^{'} \to \ell^{+}
\ell^{-}$.

\begin{figure}[t]
\epsfig{file=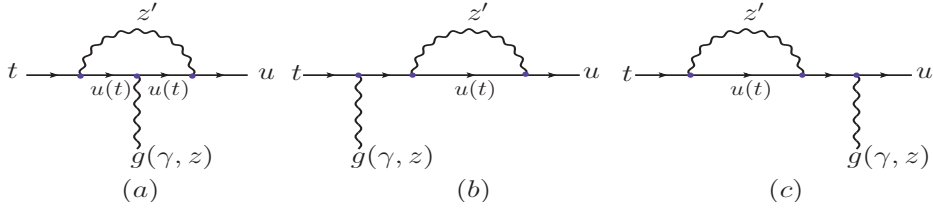,width=13cm,height=2.8cm} \vspace{-0.5cm}
\caption{The loop diagrams contributing to $t \to u V
(V=g,Z,\gamma)$.} \label{fig2}
\end{figure}
\begin{figure}[t]
\epsfig{file=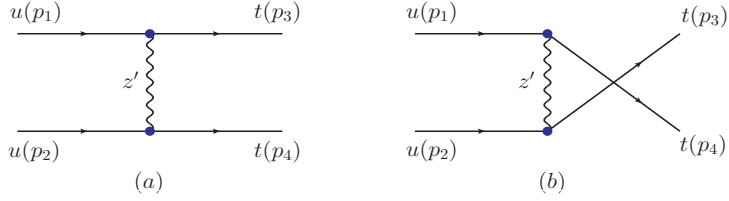,width=10cm,height=2.8cm} \vspace{-0.5cm}
\caption{Feynman diagrams contributing to $t t$ production.}
\label{fig3}
\end{figure}

Eq.(\ref{neucoup}) indicates that, unlike model-I, model-II
gives an interaction $Z^\prime \bar{t} t$ larger than $Z^\prime \bar{t} u$.
Although both models utilize the $Z^\prime \bar{t} u$ interaction
to explain the deviation of $A_{FB}^t$, model-II will have more side
effects than model-I, e.g., it may also alter sizably $R_b$,
the total and the differential rates of the $t\bar{t}$ production.
The consideration of all these effects
leads to a favorable region characterized by a very large $g_Y$
($10<\cot \theta_R < 20$) and a heavy $Z^\prime$ ($500 {\rm GeV} <
m_{Z^\prime} < 800 {\rm GeV})$ \cite{top-afb-th4}.

Since both models allow for the FCNC $Z^\prime \bar{t} u$
interaction, they may predict large top quark FCNC decays $t \to u
V$ ($V=g,Z,\gamma$) and the like-sign top pair production, as shown
in Figs.\ref{fig2} and \ref{fig3}. The analytic expressions of the
loop amplitudes for the FCNC decays are given in Appendix A. Three
points should be noted here. First, in model-I the processes $ug \to
t Z^{\prime \ast} \to t t \bar{u}$ and $u \bar{u} \to Z^{\prime
\ast} Z^{\prime \ast} \to t \bar{u} t \bar{u}$ may lead to signals
similar to the $tt$ production, which, however, are suppressed by
kinematics or high-order effects. Second, in model-II the decay $t
\to u Z$ can proceed at tree level via the $Z-Z^\prime$ mixing,
while in model-I it can only occur at loop level. Third, for the
signature $ep \to et$ at the HERA\cite{tur}, in model I, the
lepton-phobic $Z^{'}$ will not contribute to this process and can
safely avoid the constraints; For the model II, due to the large
$Z^{'}$ mass, the process $ep \to et$ is less sensitive to the
coupling of $Z^{'}u\bar{t}$. Besides, as mentioned above, the small
couplings($\tan\theta$) of $Z^{'}e^{+}e^{-}$ will cancel the large
flavor-changing coupling($\cot\theta$) of $Z^{'}u\bar{t}$ and also
not give rise to the significant contribution to process $ep \to
et$.

\section{ Numerical results and discussions}

The SM parameters used in this calculations are \cite{pdg}
\begin{eqnarray}
m_t=172.5 {\rm GeV}, m_{Z}=91.19 {\rm
~GeV},~\sin^{2}\theta_W=0.2228, ~\alpha_s(m_t)=0.1095,~\alpha=1/128.
\end{eqnarray}
For new physics parameters, we scan them within the following ranges:
\small
\begin{eqnarray*}
&& {\rm Model~ I:}~~120 {\rm ~GeV} <m_{Z^{\prime}}< 170 {\rm ~GeV}, 0.05 <\epsilon_{U} < 0.1, 0<\alpha_{x}<0.05;\\
&& {\rm Model~ II:}~~500 {\rm ~GeV}<m_{Z^{\prime}}< 2000 {\rm ~GeV},
10 <\cot\theta <20, 0.1<(V^{u}_R)_{ut}<0.2, 0<\xi_Z<0.01.
\end{eqnarray*}
\normalsize

It should be noted that the contributions to the $t\bar{t}$ cross
section mainly come from the flavor-changing t-channel in Fig.1(b)
for both models. Since the suppressions of small flavor-conserving
coupling $\epsilon_{U}$ in model I and $\tan\theta$ in model II
respectively and no interference with the SM QCD process, the
s-channel $u\bar{u} \to Z^{'} \to t\bar{t}$ has a negligible effects
on the $t\bar{t}$ production for both models in our calculations. In
additional, the measurements of $t\bar{t}$ cross section at the LHC
still have large uncertainties and may not give a new constraints on
our two models\cite{lhc-ttbar}. In our scan, we require the total
cross section of the $t\bar{t}$ production and the differential
cross section in each bin of $M_{t\bar{t}}$ to be within the
$2\sigma$ regions of their experimental values at the
Tevatron\cite{cross section,mtt}. For model-II, we also consider the
constraint from the $T$ parameter at $3\sigma$ level \cite{hexg}.
For the calculation of the hadronic cross sections, we use the
parton distribution function CTEQ6L \cite{cteq} with the
renormalization scale $\mu_R$ and factorization scale $\mu_F$
setting to be $m_t$.

\subsection{Correlations between different observables}

\begin{figure}[t]
\epsfig{file=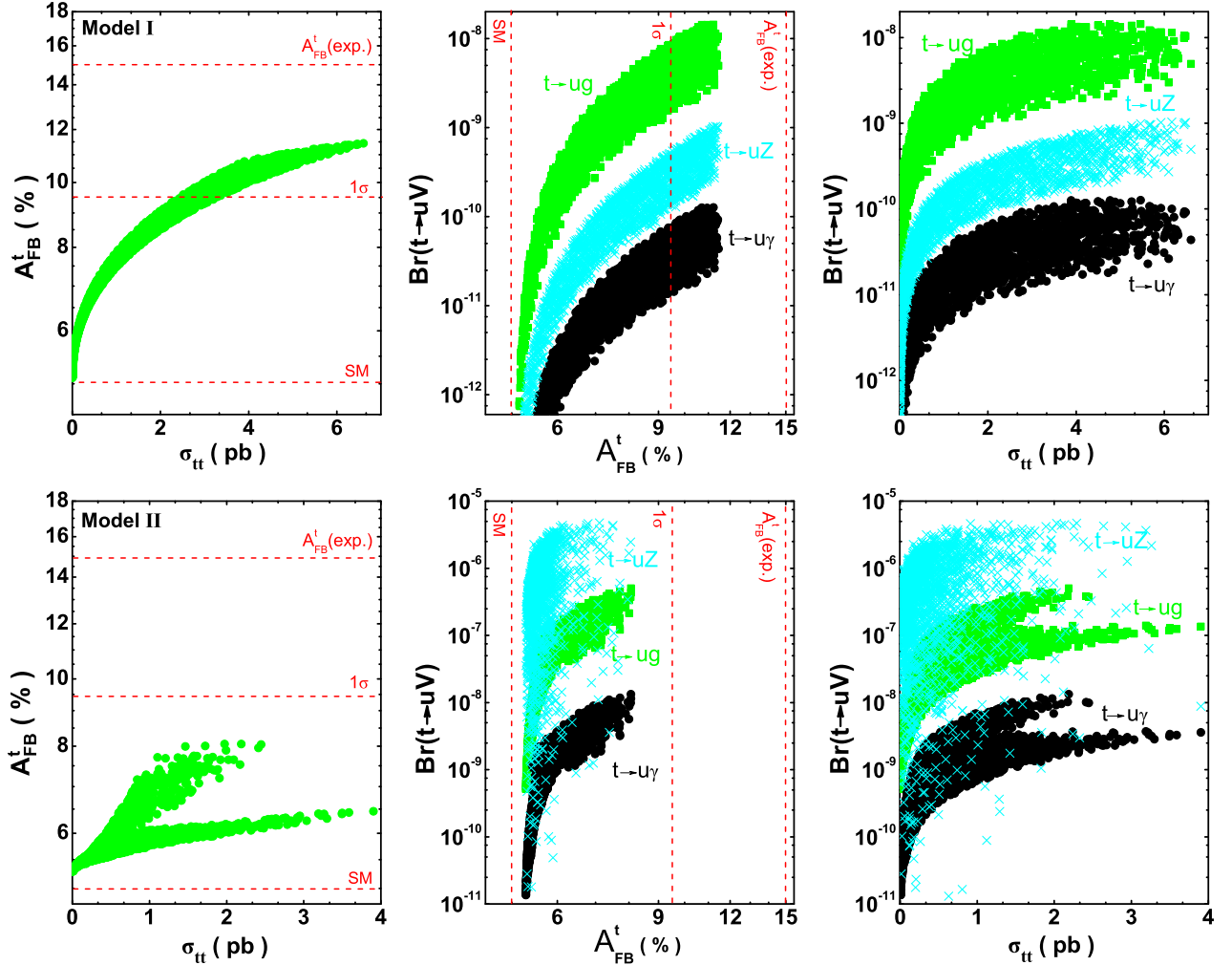,width=17cm}
\vspace*{-1.5cm}
\caption{The correlations between $A^{t}_{FB}$
at the Tevatron, the branch ratio of top FCNC decays $Br(t\to u V)$
and the cross section of the $tt$ production $\sigma(tt)$
at the LHC with $\sqrt{s}=14$ TeV.} \label{fig4}
\end{figure}

For each model we show in Fig.4 the correlations between
$A^{t}_{FB}$ at the Tevatron, the branch ratios of the top FCNC decays $Br(t\to uV)$
and the cross section of the $tt$ production at the
LHC with $\sqrt{s}=14$TeV. A common feature of the correlations is
that all these quantities are proportional to each other.
This is obvious since all the quantities receive contributions from
the same $Z^\prime \bar{t} u$ interaction.
For the forward-backward asymmetry, one can see that both models can enhance its value significantly
to alleviate the deviation, especially, model-I can reduce the deviation to $1\sigma$ level.
More details about the calculation of $A^{t}_{FB}$ were presented in \cite{top-afb-th4}.

About the FCNC decays, Fig.\ref{fig4} shows that $t \to u Z$ can
have a large branching ratio in model-II because it can proceed at
tree level via the $Z-Z^\prime$ mixing. For each decay model-II
gives a larger branching ratio than model-I because in model-I the
decays are highly suppressed by a factor $\epsilon_U^2$.
Fig.\ref{fig4} also indicates that in both models the branching
ratios for these FCNC decays are smaller than $5\times 10^{-6}$,
which are far below their current experimental bounds (namely $Br(t
\to ug)<0.02\%$ from D0 \cite{tug}, $Br(t \to u \gamma)<5.2\%$ from
ZEUS\footnote{the ZEUS collaboration gives only the upper limits of
the anomalous coupling $\kappa_{tu\gamma}$, $\kappa_{tu\gamma} <
0.174$ at 95\% CL. Using this limits and Eq.(41) in
arXiv:hep-ph/0003033 in Ref[1], we can obtain the upper limit $Br(t
\to u \gamma)<5.2\%$.} \cite{tur} and $Br(t \to u Z)<3.7\%$ from CDF
\cite{tuz}) and also smaller than the maximal values predicted in
other new physics models such as low energy supersymmetry
\cite{fcnc-susy}, Technicolor model \cite{fcnc-tc} or Little Higgs
theory \cite{fcnc-lh}. From the analysis of top FCNC decays
\cite{fcnc-susy,fcnc-tc,fcnc-lh,NLOtuv}, one can infer that
detecting these decays in the present two models would be quite
challenging at the LHC.

About the like-sign top pair production, Fig.\ref{fig4} shows that for $A_{FB}^t \geq 6\%$
the cross section in both models can be of $pb$ order, reaching $6.8 pb$
in model-I and $3.7 pb$ in model-II.
Since the signal of such a production
is characterized by two isolated like-sign leptons,
which is free from the $t\bar{t}$ background and
the huge QCD $W+$jets background \cite{like-sign top-th1,like-sign top-th2,like-sign top-th3},
it may be observable at the LHC and will be discussed in the following.
\vspace{-.3cm}

\subsection{Mass-dependent forward-backward asymmetry at the Tevatron}
\begin{figure}[t]
\vspace{-0.5cm}\epsfig{file=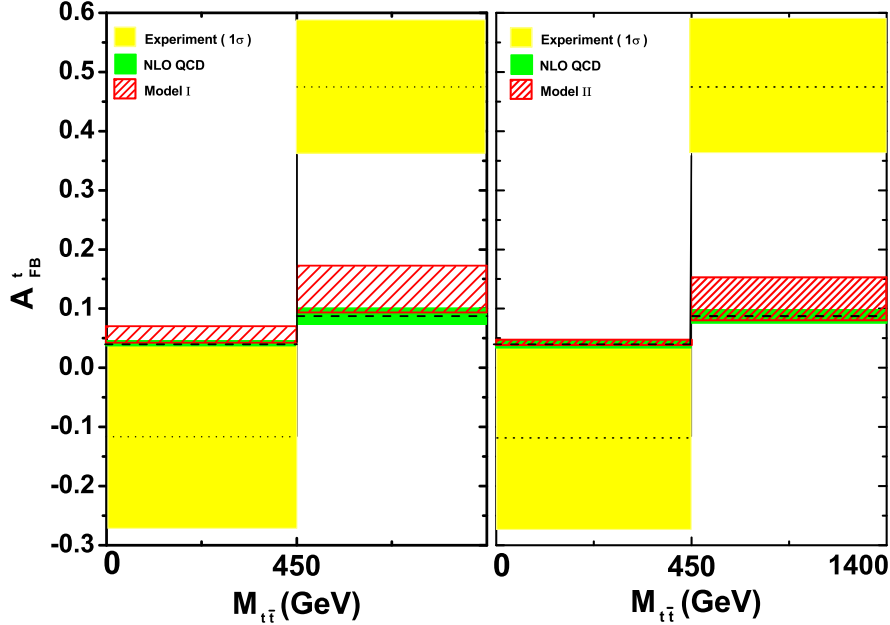,width=13cm}
\vspace{-1.0cm}
\caption{$A_{FB}^t$ at the Tevatron versus $t\bar{t}$ invariant mass.}
\label{fig5}
\end{figure}

We note that very recently the CDF reported the dependence of $A_{FB}^t$ on the
top pair invariant mass $M_{t\bar{t}}$ and found a more than $3\sigma$ discrepancy from
the SM prediction for $M_{t\bar{t}} > 450 {\rm GeV}$ \cite{top-afb-exp2}.
Motivated by this, we display the dependence of $A_{FB}^t$ on $M_{t\bar{t}}$ in Fig.\ref{fig5}.
In our calculation, we have included the SM contribution and multiplied the total
cross section by a K-factor $1.31$ to include the NLO QCD effect \cite{k-factor}.
Fig.\ref{fig5} shows that both models can enhance $A^{t}_{FB}$ in large $M_{t\bar{t}}$ region
to ameliorate the discrepancy, though the discrepancy still persists at $2\sigma$ level.
Further, we note that both models can also cause top quark polarization
asymmetry in the $t\bar t$ production at the LHC, which was studied recently
in \cite{top-afb-th2,wulei}.

\subsection{Like-sign top production at the LHC}
\begin{figure}[t]
\vspace*{-1.0cm}
\epsfig{file=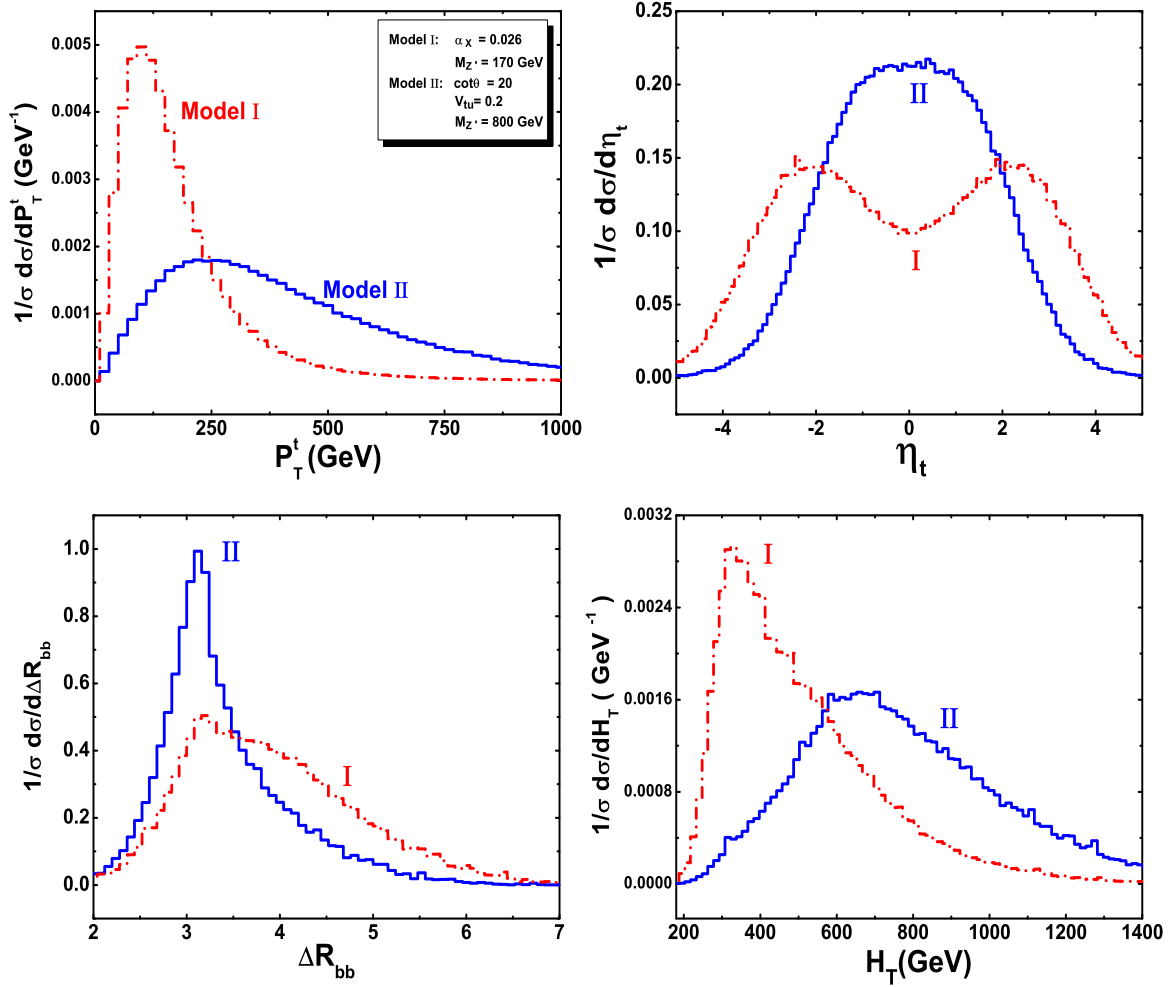,width=17cm,height=15cm}
\vspace*{-2.0cm}
\caption{The $p^{t}_{T}$, $\eta_{t}$, $\Delta R_{bb}$ and $H_{T}$
distributions for the like sign top pair production at the LHC.}
\label{fig6}
\end{figure}

Since the like-sign top production can have a large rate and its
background is low, we give further study on its observability at the
LHC. First, in Fig.\ref{fig6} we display some kinematical distributions
such as the top quark transverse momentum $p^{t}_{T}$ and its pseudo-rapidity $\eta_{t}$,
the total transverse energy $H_{T}$ of the process and the separation
between the two $b$-jets $\Delta R_{bb} \equiv \sqrt{(\Delta
\phi)^2 + (\Delta \eta)^2}$. The new physics parameters are fixed as
$\alpha_x=0.026$ and $m_{Z^\prime} =170 {\rm GeV}$ for model-I,
and $\cot \theta_R =20$ and $V_{tu}=0.2$, $m_{Z^\prime} = 800 {\rm GeV}$
for model-II.
For the last two distributions in  Fig.\ref{fig6}, we have included
the decay chain $t \to W b \to l \nu b$ in our code to simulate the signals
of the process. From the upper two and the last frames we can see that
the most events are distributed in the region with small transverse momentum
or large pseudo-rapidity. This implies that
for a light $Z^\prime$ in model-I the top quarks tend to outgo in parallel
with the beam pipe.
The third frame shows that the two $b$-jets tend
to fly in the opposite direction since they come from the back-to-back top quarks
in the $tt$ rest frame. The distinct shapes in the second frame are caused
by different masses of $Z'$ in the two models (if the $Z'$ mass is assumed to be
in the same region for both models, they give the similar shapes).

Now we discuss the detection of the $tt$ production at the LHC with $\sqrt{s}=14{\rm TeV}$.
We choose $\ell^+_i \ell^+_j b b \not{\hspace*{-0.1cm}E}_T$ ($\ell_i=e, \mu$) as the signal.
We simulate the energy resolution of the detector effects by assuming a
Gaussian smearing for the final leptons and jets \cite{enery resolution}
\begin{eqnarray}
\frac{\Delta E}{E}=\frac{5\%}{\sqrt{E}}\oplus0.55\%,~~~~{\rm for~~leptons},\\
\frac{\Delta E}{E}=\frac{100\%}{\sqrt{E}}\oplus5\%~~~~~~~~{\rm
for~~jets}~~~~~,
\end{eqnarray}
where $E$ is in GeV, and $\oplus$ indicates that the
energy-dependent and energy-independent terms are added in
quadrature. We take the $b$-jet tagging efficiency as $50\%$.
The main backgrounds are from $qq^{\prime} \to t\bar{t}W^{\pm}$ and
$qq \to W^{\pm}q^{\prime}W^{\pm}q^{\prime}$, which have been studied
in \cite{like-sign top-th2}. In our analysis we take the
same cuts as in \cite{like-sign top-th2} for the signal:
\begin{eqnarray}
&& p_{T}^\ell>15 {\rm ~GeV},
   ~E^{j}_{T}>40 {\rm ~GeV}, ~|\eta_\ell|, |\eta_j|<2.5,~
  \Delta R_{\ell j},\Delta R_{jj}>0.4, \nonumber\\
&& M(\ell_{1}j_{1}), M(\ell_{2}j_{2})<160 {\rm ~GeV}, ~M(\ell\ell jj)>500 {\rm ~GeV},
\end{eqnarray}
where $E_{T}$ denotes the transverse energy and $M$ is the invariant
mass of the final states.
Note that the production $pp \to
t\bar{t} \to bW^{+}(\to \ell^{+} \nu)\bar{b}(\to \ell^{+})W^{-}(\to
jj)$ can also mimic our signal \cite{like-sign top-th4},
which, however, has an extra jet and can be suppressed by jet veto.

Under such cuts we find the acceptance rate of the signal is
$15.5\%$ for $m_{Z^\prime} = 170 {\rm GeV}$ in model-I and $18\%$
for $m_{Z^\prime} = 800 {\rm GeV}$ in model-II.
The acceptance rate is found to increase with $m_{Z^\prime}$,
which is $12.8\%$ for $m_{Z^\prime}=120 {\rm ~GeV}$
and increased to $21\%$ for $m_{Z^\prime}=2000 {\rm ~GeV}$.
With the backgrounds
calculated in \cite{like-sign top-th2}, we get the $3\sigma$ sensitivity
for the $tt$ production by requiring $S/\sqrt{S+B} \geq 3$ for $100 fb^{-1}$
integrated luminosity. The corresponding results are listed in Table
I, where one can learn that the $tt$ production with a rate as large
as several tens of $fb$ can be detected at the LHC.

\begin{table}[t]
\caption{$3\sigma$ observation bound on the rate of the $tt$ production at the LHC with
$\sqrt{s}=14{\rm TeV}$ for $100 fb^{-1}$ integrated luminosity. \label{tab1}}
\begin{tabular}{|c|c|c|c|c|} \hline
 \multicolumn{2}{|c|}{model-I} & \multicolumn{2}{c|}{model-II} \\
\cline{1-4} ~  ~$m_{Z^{\prime}}$~&~$\sigma$~
&~$m_{Z^{\prime}}$~&~$\sigma$~ \\\hline ~ ~~ 120 {\rm GeV}~~ &
~~27.3 $fb$~~ &~~500 {\rm GeV} ~~ &~~ 19.7 $fb$~~ \\\hline ~ ~~ 150 {\rm GeV} ~~ & ~~23.6 $fb$ ~~ &~~1500 {\rm GeV} ~~ &~~
17.8 $fb$~~\\\hline ~ ~~ 170 {\rm GeV}~~  & ~~22.6 $fb$~~&~~2000 {\rm GeV}~~  &~~ 16.5 $fb$~~\\ \hline
\end{tabular}
\end{table}

\begin{figure}[t]
\epsfig{file=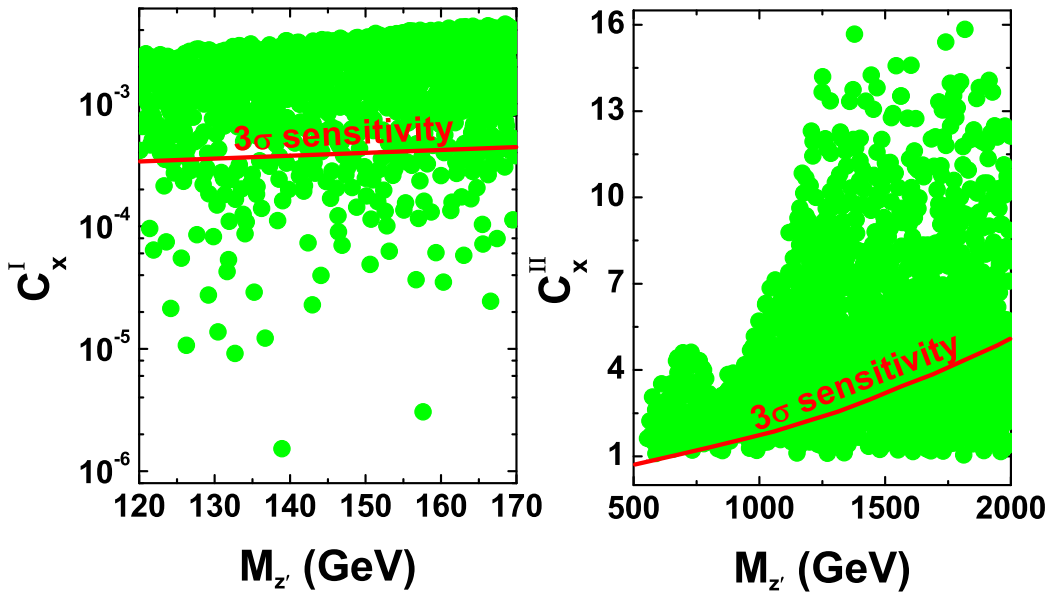,width=14cm}
\vspace*{-.7cm}
\caption{Surviving samples with
$C^I_X$ denoting $\alpha_{x}$ for model-I and
$C^{II}_X$ denoting $V_{tu}^{2}(\cot\theta+\tan\theta)^{2}$ for model-II.}
\label{fig7}
\end{figure}
Finally, in Fig.\ref{fig7} we project the surviving samples in the scan on
the  $m_{Z^\prime}-C$ plane with $C$ defined as $\alpha_X$ in model-I and
$V_{tu}^{2}(\cot\theta+\tan\theta)^{2}$ in model-II.
The solid curves in this figure are the $3 \sigma$ sensitivity and
above each curve is the observable region. Our results show
that about $83\%$ ($88\%$) of the total surviving samples
lie above the $3 \sigma$ curve for model-I (model-II).
Combined with Fig. 4, the points lying below the $3 \sigma$ curve in Fig.\ref{fig7}
give a shift of $A_{FB}^t$ less than $0.5\%$. This fact implies that if
the $tt$ production is not observed at the LHC, then our
considered models may not explain the anomaly of $A_{FB}^t$ .

\section{conclusion}
In this work we studied the correlations between $A^{t}_{FB}$ at the Tevatron,
the FCNC decays $t \to u V (V=g,Z,\gamma)$ and the like-sign
top pair production at the LHC in two models with non-universal $Z^\prime$ interactions, i.e.
the left-right model and the $\mathrm{U}(1)_X$ model.
We also studied the dependence of $A^{t}_{FB}$ on the top pair invariant mass $M_{t\bar{t}}$.
We found that under the current experimental constraints both models can
alleviate the deviation of $A^{t}_{FB}$ and, meanwhile, enhance the $tt$ production
to the detectable level of the LHC. We also
found that, since the two models give different predictions for the
observables and also their correlations, they may be distinguished
by jointly studying these observables.
In particular, we emphasize that exploring
the $tt$ production at the LHC will allow for a further test of the models
which are used to explain the anomaly of $A_{FB}^t$ observed at the Tevatron.

Note Added : Months after our manuscript finished, the CMS
collaboration reported their result of searching the like-sign top
pair induced by the FCNC $Z^{'}$ at the LHC. The limit of the cross
section $\sigma(pp\to tt(j)) <17.0$ pb at 95 \% CL\cite{cms-tt}. We
found that in our parameter space, the like-sign top pair cross
sections are far bellow the upper limits set by the CMS.

\section*{Acknowledgement}
This work was supported in part by HASTIT under grant No.
2009HASTIT004, by the National Natural Science Foundation of China
(NNSFC) under grant Nos. 10821504, 10725526, 10775039,
11075045 and by the Project of Knowledge Innovation Program (PKIP)
of Chinese Academy of Sciences under grant No. KJCX2.YW.W10.

\appendix
\section{Analytic expressions for FCNC decay amplitudes }
Here we list the amplitudes for the loop induced processes $t\to u
V(V=g,\gamma,Z)$ shown in Fig.2. The notation $\mathcal{M}(q)$
denotes the amplitude of the diagram with  quark $q$ ($q=u,t$)
appeared in the loop.

For model-I, the  amplitudes are given by \small
\begin{eqnarray}
\mathcal{M}^{(a)}_{g}(q)&=&ag_{s}T^{A}_{\beta\alpha}\bar{u}(p_{u})
[-4C^{\nu\rho}\gamma_{\rho}-\gamma^{\nu}+2B_{0}^{1}\gamma^{\nu}-2(\slashed
p_{t}-\slashed p_{u})\gamma^{\nu}\gamma^{\rho}
C_{\rho}]P_{R}u(p_{t})\varepsilon^{\ast}_{\nu}(p_{g}),
\\
\mathcal {M}^{(b)}_{g}(q)&=&ag_{s}T^{A}_{\beta\alpha}\bar{u}(p_{u})
\frac{(2B^{2}_{1}+1)m_{t}\slashed
p_{u}\gamma^{\nu}P_{L}}{p_{u}^{2}-m_{t}^{2}}u(p_{t})\varepsilon^{\ast}_{\nu}(p_{g}),
\\
\mathcal {M}^{(c)}_{g}(q)&=&ag_{s}T^{A}_{\beta\alpha}\bar{u}(p_{u})
\frac{(2B_{1}^{3}+1)m_{t}^{2}\gamma^{\nu}P_{R}}{p_{t}^{2}-m_{u}^{2}}u(p_{t})\varepsilon^{\ast}_{\nu}(p_{g}),
\\
\mathcal {M}^{(a)}_{\gamma}(q)&=&\frac{2}{3}ea\bar{u}(p_{u})
[-4C^{\nu\rho}\gamma_{\rho}-\gamma^{\nu}+2B_{0}^{1}\gamma^{\nu}
-2(\slashed p_{t}-\slashed p_{u})\gamma^{\nu}\gamma^\rho
C_{\rho}]P_{R}u(p_{t})\varepsilon^{\ast}_{\nu}(p_{\gamma}),
\\
\mathcal {M}^{(b)}_{\gamma}(q)&=&\frac{2}{3}ea\bar{u}(p_{u})
\frac{(2B^{2}_{1}+1)m_{t}\slashed
p_{u}\gamma^{\nu}P_{L}}{p_{u}^{2}-m_{t}^{2}}u(p_{t})\varepsilon^{\ast}_{\nu}(p_{\gamma}),
\\
\mathcal {M}^{(c)}_{\gamma}(q)&=&\frac{2}{3}ea\bar{u}(p_{u})
\frac{(2B_{1}^{3}+1)m_{t}^{2}\gamma^{\nu}P_{R}}{p_{t}^{2}-m_{u}^{2}}u(p_{t})\varepsilon^{\ast}_{\nu}(p_{\gamma}),
\\
\mathcal
{M}^{(a)}_{Z}(u)&=&-a\delta_{\alpha\beta}g\bar{u}(p_{u})(\frac{2}{3}
\sin\theta_{W}\tan\theta_{W})
[-4C^{\nu\rho}\gamma_{\rho}-\gamma^{\nu}+2B_{0}^{1}\gamma^{\nu}\nonumber
\\&&-2(\slashed
p_{t}-\slashed p_{u})\gamma^{\nu}\gamma^{\rho}
C_{\rho}]P_{R}u(p_{t})\varepsilon^{\ast}_{\nu}(p_{Z}),
\\
\mathcal {M}^{(a)}_{Z}(t)&=&-a\delta_{\alpha\beta}g\bar{u}(p_{u})
\biggl[\frac{m_{t}^{2}}{\cos\theta_{W}}C_{0}\gamma^{\nu}P_{R}+\frac{2}{3}
\sin\theta_{W}\tan\theta_{W}[-4C^{\nu\rho}\gamma_{\rho}\nonumber
\\&&-\gamma^{\nu}+2B_{0}^{1}\gamma^{\nu} -2(\slashed p_{t}-\slashed
p_{u})\gamma^{\nu}\gamma^{\rho} C_{\rho}]P_{R}\biggl]
u(p_{t})\varepsilon^{\ast}_{\nu}(p_{Z}),
\\
\mathcal {M}^{(b)}_{Z}(q)&=&-a\delta_{\alpha\beta}g\bar{u}(p_{u})
(\frac{4\sin^{2}\theta_{W}-3}{6\cos\theta_{W}})
\biggl[\frac{(2B^{2}_{1}+1)m_{t}\slashed
p_{u}\gamma^{\nu}P_{L}}{p_{u}^{2}-m_{t}^{2}}\biggl]u(p_{t})\varepsilon^{\ast}_{\nu}(p_{Z}),
\\
\mathcal {M}^{(c)}_{Z}(q)&=&-a\delta_{\alpha\beta}g\bar{u}(p_{u})
(\frac{2}{3}
\sin\theta_{W}\tan\theta_{W})\frac{(2B_{1}^{3}+1)m_{t}^{2}\gamma^{\nu}P_{R}}{p_{t}^{2}-m_{u}^{2}}
u(p_{t})\varepsilon^{\ast}_{\nu}(p_{Z}).
\end{eqnarray}

For model-II,  the amplitudes are given by
\begin{eqnarray}
\mathcal {M}^{(a)}_{g}(u)&=&b g_{s}T_{\beta\alpha}^{A}\bar{u}(p_{u})
[-4C^{\nu\rho}\gamma_{\rho}-\gamma^{\nu}+2B_{0}^{1}\gamma^{\nu}-
2(\slashed p_{t}-\slashed p_{u})\gamma^{\nu}\gamma^{\rho}
C_{\rho}]g^{u}_{Z^{\prime}R}P_{R}u(p_{t})\varepsilon^{\ast}_{\nu}(p_{g}),
\\
\mathcal {M}^{(a)}_{g}(t)&=&
bg_{s}T_{\beta\alpha}^{A}
\bar{u}(p_{u})\{[-4C^{\nu\rho}\gamma_{\rho}-\gamma^{\nu}+2B_{0}^{1}\gamma^{\nu}-
2(\slashed p_{t}-\slashed p_{u})\gamma^{\nu}\gamma^{\rho}
C_{\rho}]g^{t}_{Z^{\prime}R}P_{R}\nonumber
\\&&+8m_{t}C^{\nu}g^{t}_{Z^{\prime}L}P_{L}-4m_{t}(p_{u}-p_{t})^{\nu}C_{0}
g^{t}_{Z^{\prime}L}P_{L}\}u(p_{t})\varepsilon^{\ast}_{\nu}(p_{g}),
\\
\mathcal {M}^{(b)}_{g}(u)&=&bg_{s}T_{\beta\alpha}^{A}
\bar{u}(p_{u}) \frac{(2B^{2}_{1}+1)m_{t}\slashed
p_{u}\gamma^{\nu}g^{u}_{Z^{\prime}R}P_{L}}{p_{u}^{2}-m_{t}^{2}}u(p_{t})\varepsilon^{\ast}_{\nu}(p_{g}),
\\
\mathcal {M}^{(b)}_{g}(t)&=&bg_{s}T_{\beta\alpha}^{A}\bar{u}(p_{u})
\{(2B^{2}_{1}+1)m_{t}\slashed
p_{u}\gamma^{\nu}g^{t}_{Z^{\prime}R}P_{L}+(4B^{2}_{0}-2)m_{t}\slashed
p_{u}\gamma^{\nu}g^{t}_{Z^{\prime}L}P_{L}\nonumber\\
&& +(4B^{2}_{0}-2)m_{t}^{2} \gamma^{\nu}g^{t}_{Z^{\prime}L}P_{R}\}
\frac{u(p_{t})\varepsilon^{\ast}_{\nu}(p_{g})}{p_{u}^{2}-m_{t}^{2}},
\end{eqnarray}
\begin{eqnarray}
\mathcal {M}^{(c)}_{g}(u)&=&bg_{s}T_{\beta\alpha}^{A}
\bar{u}(p_{u})
\frac{(2B_{1}^{3}+1)m_{t}^{2}\gamma^{\nu}g^{u}_{Z^{\prime}R}P_{R}}{p_{t}^{2}-m_{u}^{2}}u(p_{t})\varepsilon^{\ast}_{\nu}(p_{g}),
\\
\mathcal {M}^{(c)}_{g}(t)&=&bg_{s}T_{\beta\alpha}^{A}
\bar{u}(p_{u})
\frac{(2B_{1}^{3}+1)m_{t}^{2}\gamma^{\nu}g^{t}_{Z^{\prime}R}P_{R}+(4B_{0}^{6}-2)m_{t}\gamma^{\nu}\slashed
p_{t}g^{t}_{Z^{\prime}L}P_{L}}
{p_{t}^{2}-m_{u}^{2}}u(p_{t})\varepsilon^{\ast}_{\nu}(p_{g}),
\\
\mathcal {M}^{(a)}_{\gamma}(u)&=&
\frac{2}{3}eb\bar{u}(p_{u})
[-4C^{\nu\rho}\gamma_{\rho}-\gamma^{\nu}+2B_{0}^{1}\gamma^{\nu}-
2(\slashed p_{t}-\slashed p_{u})\gamma^{\nu}\gamma^{\rho}
C_{\rho}]g^{u}_{Z^{\prime}R}P_{R}u(p_{t})\varepsilon^{\ast}_{\nu}(p_{\gamma}),
\\
\mathcal {M}^{(a)}_{\gamma}(t)&=&
\frac{2}{3}eb
\bar{u}(p_{u})\{[-4C^{\nu\rho}\gamma_{\rho}-\gamma^{\nu}+2B_{0}^{1}\gamma^{\nu}-
2(\slashed p_{t}-\slashed p_{u})\gamma^{\nu}\gamma^{\rho}
C_{\rho}]g^{t}_{Z^{\prime}R}P_{R}\nonumber
\\&&+8m_{t}C^{\nu}g^{t}_{Z^{\prime}L}P_{L}-4m_{t}(p_{u}-p_{t})^{\nu}C_{0}
g^{t}_{Z^{\prime}L}P_{L}\}u(p_{t})\varepsilon^{\ast}_{\nu}(p_{\gamma}),
\\
\mathcal {M}^{(b)}_{\gamma}(u)&=&
\frac{2}{3}eb\bar{u}(p_{u}) \frac{(2B^{2}_{1}+1)m_{t}\slashed
p_{u}\gamma^{\nu}g^{u}_{Z^{\prime}R}P_{L}}{p_{u}^{2}-m_{t}^{2}}u(p_{t})\varepsilon^{\ast}_{\nu}(p_{\gamma}),
\\
\mathcal {M}^{(b)}_{\gamma}(t)&=&
\frac{2}{3}eb\bar{u}(p_{u})\{(2B^{2}_{1}+1)m_{t}\slashed
p_{u}\gamma^{\nu}g^{t}_{Z^{\prime}R}P_{L}+(4B^{2}_{0}-2)m_{t}\slashed
p_{u}\gamma^{\nu}g^{t}_{Z^{\prime}L}P_{L} \nonumber \\
&&+(4B^{2}_{0}-2)m_{t}^{2} \gamma^{\nu}g^{t}_{Z^{\prime}L}P_{R}\}
\frac{u(p_{t})\varepsilon^{\ast}_{\nu}(p_{\gamma})}{p_{u}^{2}-m_{t}^{2}},
\\
\mathcal
{M}^{(c)}_{\gamma}(u)&=&\frac{2}{3}eb\bar{u}(p_{u})
\frac{(2B_{1}^{3}+1)m_{t}^{2}\gamma^{\nu}g^{u}_{Z^{\prime}R}P_{R}}{p_{t}^{2}-m_{u}^{2}}u(p_{t})\varepsilon^{\ast}_{\nu}(p_{\gamma}),
\\
\mathcal {M}^{(c)}_{\gamma}(t)&=&\frac{2}{3}eb\bar{u}(p_{u})
\frac{(2B_{1}^{3}+1)m_{t}^{2}\gamma^{\nu}g^{t}_{Z^{\prime}R}P_{R}+(4B_{0}^{3}-2)m_{t}\gamma^{\nu}\slashed
p_{t}g^{t}_{Z^{\prime}L}P_{L}}
{p_{t}^{2}-m_{u}^{2}}u(p_{t})\varepsilon^{\ast}_{\nu}(p_{\gamma}).
\end{eqnarray}
In above expressions, $p_t$ and $p_u$ denote the momenta of the top and up quark
respectively, $B$ and $C$ are loop functions defined in \cite{loop}
and calculated by LoopTools \cite{Hahn}, the dependence of the loop
functions on momentums and masses is given by
\begin{eqnarray}
B^{1}(q)&=&B(p_{t},m_{Z^{\prime}},m_{q}), \quad \quad B^{2}(q)=B(-p_{u},m_{q},m_{Z^{\prime}}), \nonumber \\
B^{3}(q)&=&B(-p_{t},m_{q},m_{Z^{\prime}}), \quad \quad
C(q)=C(-p_{u},p_{t},m_{q},m_{Z^{\prime}},m_{q}), \nonumber
\end{eqnarray}
and the constants are defined by
\small
\begin{eqnarray}
a&=&-\frac{i}{16\pi^{2}}\epsilon_{\mu}g^{2}_{x}, \label{supfactor}\\
b&=&\frac{i}{16\pi^{2}}\frac{e^{2}V^{u\ast
}_{Rtu}V^{u}_{Rtt}}{4\cos^{2}\theta_{W}\sin\theta_{W}}
(\tan\theta_{R}+\cot\theta_{R}),\\
g^{u,t}_{Z^{\prime}L}&=&(1-\frac{4}{3}\sin^{2}\theta_{W})\xi+\frac{1}{3}\sin\theta_{W}\tan\theta_{R},\\
g^{u}_{Z^{\prime}R}&=&-\frac{4}{3}\sin^{2}\theta_{W}\xi+\frac{4}{3}\sin\theta_{W}\tan\theta_{R} - \sin \theta_W (\tan \theta_R + \cot \theta_R ) V_{Rtu}^{u \ast} V_{Rtu}^u ,\\
g^{t}_{Z^{\prime}R}&=& -\frac{4}{3}\sin^{2}\theta_{W}\xi
+\frac{1}{3}\sin\theta_{W}\tan\theta_{R}- \sin\theta_{W} (\tan
\theta_R + \cot\theta_{R}) V_{Rtt}^{u \ast} V_{Rtt}^u.
\end{eqnarray}

\end{document}